\documentclass{aastex63}

\graphicspath{{./}{figures/}}

\received{September 23, 2019}
\revised{November 04, 2019}
\accepted{November 15, 2019}
\submitjournal{ApJL}

\shorttitle{Long lived dust rings around HD169142}
\shortauthors{C. Toci et al.}
\graphicspath{{./}{figures/}}

\begin{document}

\title{Long lived dust rings around HD169142}

\correspondingauthor{Claudia Toci}
\email{claudia.toci@inaf.it}

\author[0000-0002-6958-4986]{Claudia Toci}
\affiliation{INAF - Osservatorio Astronomico di Brera, \\
Via Brera 28, 20121 Milan, Italy}

\author[0000-0002-2357-7692]{Giuseppe Lodato}
\affiliation{Dipartimento di Fisica, Universit\'a degli Studi di Milano, \\
Via Giovanni Celoria 16, I-20133 Milano, Italy}

\author[0000-0001-6156-0034]{Davide Fedele}
\affiliation{INAF - Osservatorio Astrofisico di Arcetri, \\
Largo E. Fermi 5, I-50125, Florence, Italy}

\author[0000-0003-1859-3070]{Leonardo Testi}
\affiliation{European Southern Observatory \\ Karl-Schwarzschild-Str 2, D-85748 Garchin, Germany}
\affiliation{INAF - Osservatorio Astrofisico di Arcetri, \\
Largo E. Fermi 5, I-50125, Florence, Italy}
\affiliation{Excellence Cluster ‘ORIGINS’, \\ Boltzmann-Str. 2, D-85748 Garching, Germany}

\author[0000-0001-5907-5179]{Christophe Pinte}
\affiliation{Monash Centre for Astrophysics (MoCA) and School of Physics and Astronomy, \\ Monash University, Clayton Vic 3800, Australia}



\begin{abstract}
Recent ALMA observations of the protoplanetary disc around HD~169142 reveal a peculiar structure made of concentric dusty rings: a main ring at $\sim$~20~au, a triple system of rings at  $\sim 55-75$~au in millimetric continuum emission and a 
perturbed gas surface density from the $^{12}$CO,$^{13}$CO and C$^{18}$O~$(J=2-1)$ surface brightness profile.
In this Letter, we perform three-dimensional numerical simulations and radiative transfer modeling exploring the possibility that two giant planets interacting with the disc and orbiting in resonant locking can be responsible for the origin of the observed dust inner rings structure. We find that in this configuration the dust structure is actually long lived while the gas mass of the disc is accreted onto the star and the giant planets, emptying the inner region. In addition, we also find that the innermost planet is located at the inner edge of the dust ring, and can accrete mass from the disc, generating a signature in the dust ring shape that can be observed  in mm ALMA observations. 
\end{abstract}

%
%
\section{Introduction} \label{sec:intro}
The initial conditions for the formation of planets naturally determine the characteristics of the resulting planetary system \citep{Mordasini2016} and are connected to the properties of the protoplanetary discs where planet formation occurs. 
Several substructures within these discs have been imaged in great detail with ALMA and VLT/SPHERE (e.g, \citealt{Andrews2018,Benisty2015}). 
The most common kind of substructure appear to be in the form of gaps and rings, which play an important role due to their possible connection with ongoing or subsequent formation of planets.
Rings seem indeed to be common and they are found at any distance to the star, with no correlation between their location and the host star luminosity \citep{Long2018,Huang2018}, but occur with different morphologies \citep{Liu2019}. It it still under debate whether these structures are long lived and thus represent a feature that is related to an evolutionary stage of the system or they are transient structures, created and destroyed on shorter timescale compared with the disc lifetime. This question is closely related to the estimate of the age of protoplanetary discs and has important implications on the dust and planetesimal evolution: a long-lived ring can be an ideal environment for grain growth \citep{Brauer2008}. 
Naturally, their lifetimes are dependent on the mechanism that generates them. 
Indeed, in order to explain their presence many theoretical models have been proposed, including dead zones edges \citep{Pinilla2016} self-induced reconnection in magnetized discwind systems \citep{Suriano2018}, photoevaporation  \citep{Alexander2014} or dust processes \citep{Birnstiel2015,Gonzalez2015}.
Another possibility is that the presence of planets embedded in the discs create pressure bumps in the gas, confining the dust in rings as discussed in \citet{Pinilla2012}.
The recent discoveries of planets inside gaps, either by direct imaging \citep{Keppler2019} or by their kinematic signatures \citep{Pinte2019} and the fact that the location of gaps does not coincide with the expected snow lines \citep{Huang2018,Long2018} lends credits to the idea that at least some of these gaps mark the presence of young planets \citep{ALMA2015,Dipierro2015,Lodato2019}.
This is supported by the comparison of the direct and indirect detection of candidate protoplanets with synthetic images obtained performing radiative transfer and hydrodynamical simulations (see i.e \citealt{Dipierro2015}).
According to current theories, a planet still embedded in its disc should accrete mass and open a partially depleted gap in the circumstellar disk \citep{Crida2006}. The accreted material is expected to orbit around the planet, forming a rotating
circumplanetary disc.
Up to date, a few candidates still embedded
in their parental circumstellar discs have been claimed  \citep{Isella2019},  but most still lack confirmation \citep{Testi2015}.

The young Herbig Ae/Be star HD 169142 has an age of $6^{+6}_{-3}$ Myr and a mass of M$_{\star} = 1.65$~M$_{\odot}$. Its distance is 117 $\pm$ 4~pc \citep{Gaia2016}.
According to \citet{Panic2008} observations in the 1.3 mm dust continuum and CO $(J = 2-1)$ emission line the disc has an inclination of 13 degrees, a position angle of 5 degrees and a total gas mass of  $\sim 0.6-3.0 \times 10^{-2}$~M$_\odot$, confirmed by subsequent observations  (\citealt{Fedele2017,Perez2019}, hereafter FD17 and PS19).
This source has been observed in the thermal mid-infrared emission \citep{Honda2012}, near-infrared emission \citep{Reggiani2014,Pohl2017,Ligi2018,Bertrang2018}, near-infrared polarimetric and scattered light images \citep{Pohl2017}. 
Gas and dust are physically decoupled in HD~169142. The source shows two dust rings in continuum emission: an inner one between $\sim 20-35$~au, heavily depleted of dust and gas (FD17; \citealt{Macias2018}, hereafter ME18) and an outer one at $\sim 55-85$~au, recently resolved as a system of three thin rings by observations of PS19, while CO isotopologous emission maps of FD17 point out that the gas has an inner cavity ( $\sim 50$~au), with surface density reduced by a factor of $\sim 30-40$ (FD17, ME18).

Many authors (FD17, \citealt{Pohl2017,Ligi2018,Bertrang2018}, ME18, PS19) pointed out the presence of giant planets in the disc (M~$>$ few Jupiter masses, hereafter M$_J$), an inner one located inside the inner dust cavity ($\rm{R} < 20$~au) and an outer one in the gap between the two dust rings ($35 < \rm{R} < 55$~ au) to explain the signature double-ring morphology. The splitting of the outer ring in concentric rings have been modeled by PS19 with the presence of a single migrating low mass planet ($\sim 10^{-2}$~ M$_J$).
Potential signatures due to the presence of a candidate massive protoplanet could have been detected in this source via direct imaging in the radio/near-infrared \citep{Reggiani2014,Ligi2018,Gratton2019}, but the nature of the candidates still have to be confirmed.

In this letter we model the two prominent dust rings and the cavity of HD~169142 including in the system two giant planets, in order to check whether this substructure is a transient or is actually a long lived feature, giving constrains on the ratio between the masses and the position of the planets. We present the results assuming a single value for the gas and dust mass of the protoplanetary disc, showing that we can obtain a long lived dust ring trapped between two giant planets planets with resonant orbits. In a fortcoming paper we will present a more general study on this mechanism.

\section{Hydrodynamical simulation} \label{sec:methods}
We perform 3D global simulations of dust and gas disc with two embedded giant protoplanets using the smoothed particle hydrodynamics (SPH) code PHANTOM (see \citealt{Lodato2010,Price2017}).
We assume a single grain size of 1.00~mm, employ the  single-fluid algorithm based on the terminal velocity approximation \citep{Laibe2014}. The conservation of the dust mass is granted due to the prescription of \cite{Ballabio2018}. 

We consider a star with mass 1.65 M$_\odot$ and two planets allowed to migrate and accrete mass. 
They are initialized at a location of 18 and 55 au and with masses of $\rm{M}_{\rm{1}}$=1.5 M$_J$ and $\rm{M}_{\rm{2}}$=0.5 M$_J$ respectively, so that $\rm{M}_{\rm{2}}$/ $\rm{M}_{\rm{1}} < 1$ initially. 
They are massive enough to open a gap in the
gas surface density according to \cite{Crida2006} but are initially smaller in masses and further out in position with respect to previous models (see i.e \citealt{Pohl2017}), in order to account for the accretion of dust and gas and the migration of planets during the simulation. Our time unit is the period of the outer planet at its initial position (55 au, $\rm{T}_{orb}\sim 320$~yr).

We represent the protoplanetary bodies  and the central star  using sink particles \citep{Bate1995} with accretion radii of 1/8 of the Hill radius \citep{Dipierro2015} for the planets (0.14 and 0.4~au respectively). For the central star, due to the fact that our aim is to model the gas and dust dynamics at $\sim 20$~au, we chose an accretion radius of 5~au, not considering the innermost part of the disc behaviour. We also tested our results for an accretion radius of 1~au, finding no significant differences. We choose not to include a third body (as predicted by PS19), used to model also the outer ring splitting. As already stated, our aim is to characterise the inner structure as a function of the two giant planets.
Dust and gas can be accreted onto the sinks, representing a condition that mimics a real accretion scenario and the planets are free to migrate due to planet-disc interaction, planet-planet interaction and viscous disc evolution. Thus, in our simulations, all the material that enters the sink radius is considered as instantaneously accreted. This may over-estimate the accretion onto the planet, and we remark that in the following, when we refer to the ``planet mass'', this should actually be considered as the mass of the planet and of its circumplanetary disc. 

We set up the discs in PHANTOM using the setup from 
\cite{Lodato2010}. The scenario is modeled with the central star  surrounded by a disc of $10^6$ SPH particles,
extending from $\rm{R}_{\rm{in}} = 5$ to $\rm{R}_{\rm{out}} = 150$~au. The surface density profile $\Sigma(\rm{R})$ is 
\begin{equation}
\Sigma(\rm{R})= \Sigma_0\left(\frac{\rm{R}}{\rm{R}_0}\right)^{-p}\exp{\left[ -(\rm{R}/\rm{R}_0)^{(2-q)}\right]}\left(1-\sqrt{\frac{\rm{R}_{\rm{in}}}{\rm{R}}}\right),   
\end{equation}
where $\rm{R}_0=100$ au, $p$ and $q$ are two indexes and the density normalization $\Sigma_0=3.6$ g~cm$^{-2}$ is set to have an initial total gas mass of $\sim 10^{-2}~\rm{M}_\odot$.
We assume a locally isothermal equation of state where the sound speed profile is $c_s = c_{s,\rm{ref}}(\rm{R}/\rm{R}_{\rm{ref}})^{-q}$. The aspect ratio in the gas is given by
$\rm{H/R} = (\rm{H/R})_{\rm{ref}} (\rm{R/R}_{\rm{ref}})^{(1/2-q)}$. We also assumed an aspect ratio value of $\rm{H/R}=0.07$ at $\rm{R=R}_{\rm{ref}}$ and $p=q=0.5$ as found in FD17.
We set the SPH $\alpha_{AV}$ viscosity to obtain an effective \citet{SS1973} viscosity $\alpha_{SS} = 0.005$.

The initial dust mass is $1.2 \times 10^{-4} \rm{M}_\odot$, close to the value found by FD17 based on the 1.3 continuum observation. 
The dust surface density has initially the same functional form of the gas, but is scaled down assuming a gas-to-dust ratio of 80 as inferred from FD17 CO isotopologous measurements. 
For our chosen dust size of $a = 1$ mm (and assuming an internal density of the dust of $\rho_d = 3$~g cm$^{-2}$), the initial Stokes number, defined as $\sim S_t=\rho_d a/\Sigma_{gas} $, is smaller than 1. This implies that dust and the gas are initially coupled in the disc.

\subsection{Radiative transfer model}
To directly compare \citep{Pinte2006} with observations of HD~169142 we
post-process our simulation using the Monte Carlo
radiative transfer code MCFOST \citep{Pinte2009}. The code directly maps from the SPH particle to the radiative transfer grid physical quantities such as density and temperature thanks to a Voronoi tesselation that matches each cell with an SPH particle, without any interpolation. The astronomical silicates and carbons are assumed from \cite{Draine2003}.
The grain size distribution has a slope of $-3.5$ and goes from 0.03~$\mu$m to 3~mm. To irradiate the disc the star is treated as an isotropic source with stellar spectral models obtained from the 5Myr isochrones from \cite{Siess2000}, with $\rm{T}_{eff}\sim$ ~8000~K and a radius $R\sim 2.2~\rm{R}_\odot$.
The source is located at the distance of 117~pc, adopting the disc inclination and position angle of 13 and 5 degrees.

\section{Results} \label{sec:results}

\begin{figure}[ht!]
\centering
\includegraphics[width=0.6\linewidth]{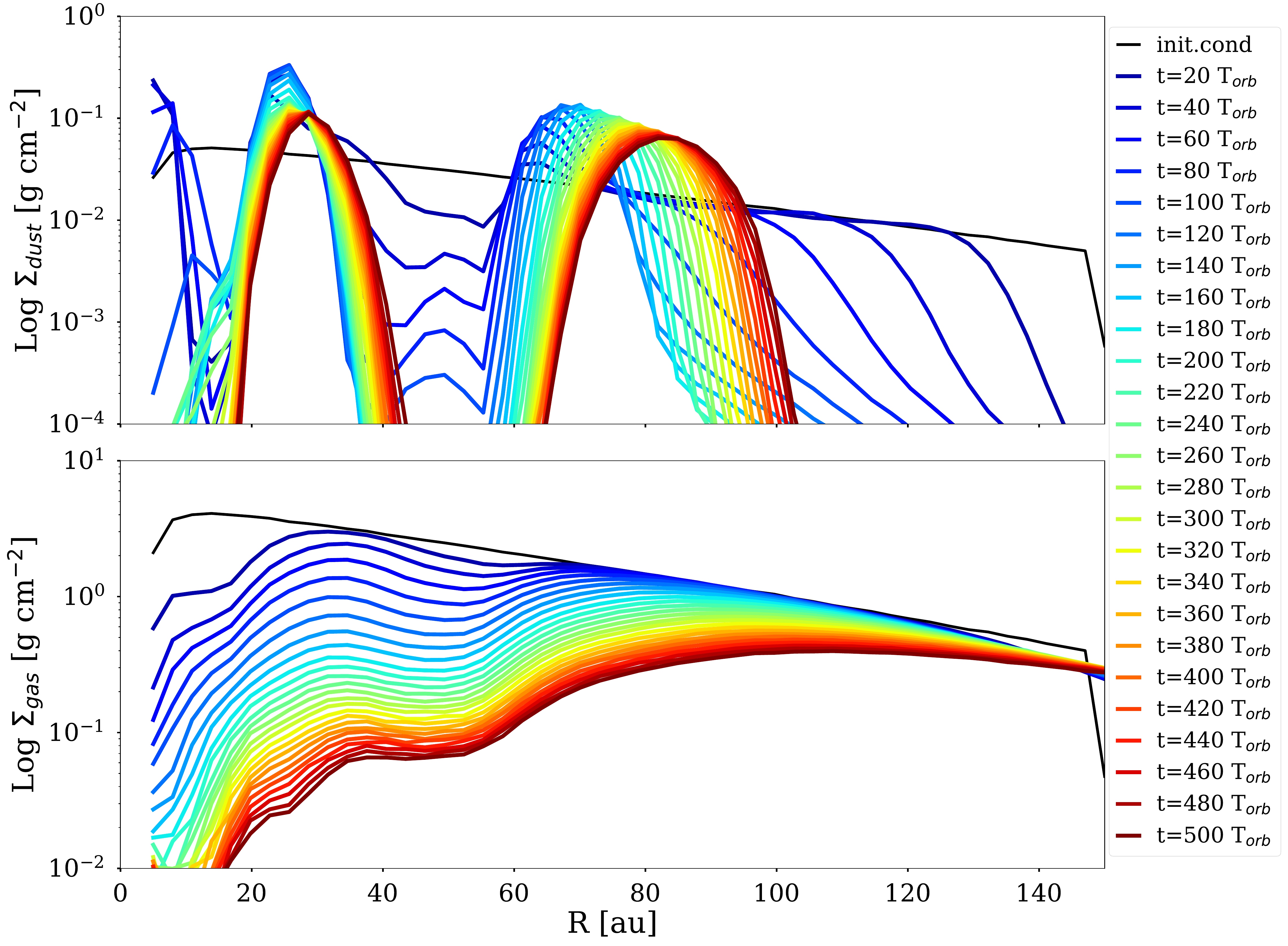}
\caption{ $\Sigma_{dust}$  (top) and $\Sigma_{gas}$ (bottom) as a function of the radius R for increasing time. In black the initial surface density profiles.  {Our time unit is normalised} with the initial orbital period of the outer planet ($\sim$ 320~yr).}
\label{fig:1}
\end{figure}
Our reference profile for the gas and dust distributions is the DALI thermochemical model from FD17, that models the dust surface density profile $\Sigma_{dust}$ with two dust rings ($20-35$~au and $56-83$~au) and the gas density profiles $\Sigma_{gas}$ with an inner cavity ($13-56$~au) reduced by a factor 30 - 40.
However, we considered an update of the published model of FD17, limited by the spatial resolution of the available data at the time.
Based on the $^{13}$CO J=3-2 ALMA archive data from \cite{Honda2012}, 
 the gas cavity appears to be located at $\sim 20-56$ au. 
\begin{figure}[ht!]
\centering
\begin{minipage}[c]{0.45\linewidth}
\includegraphics[width=1\linewidth]{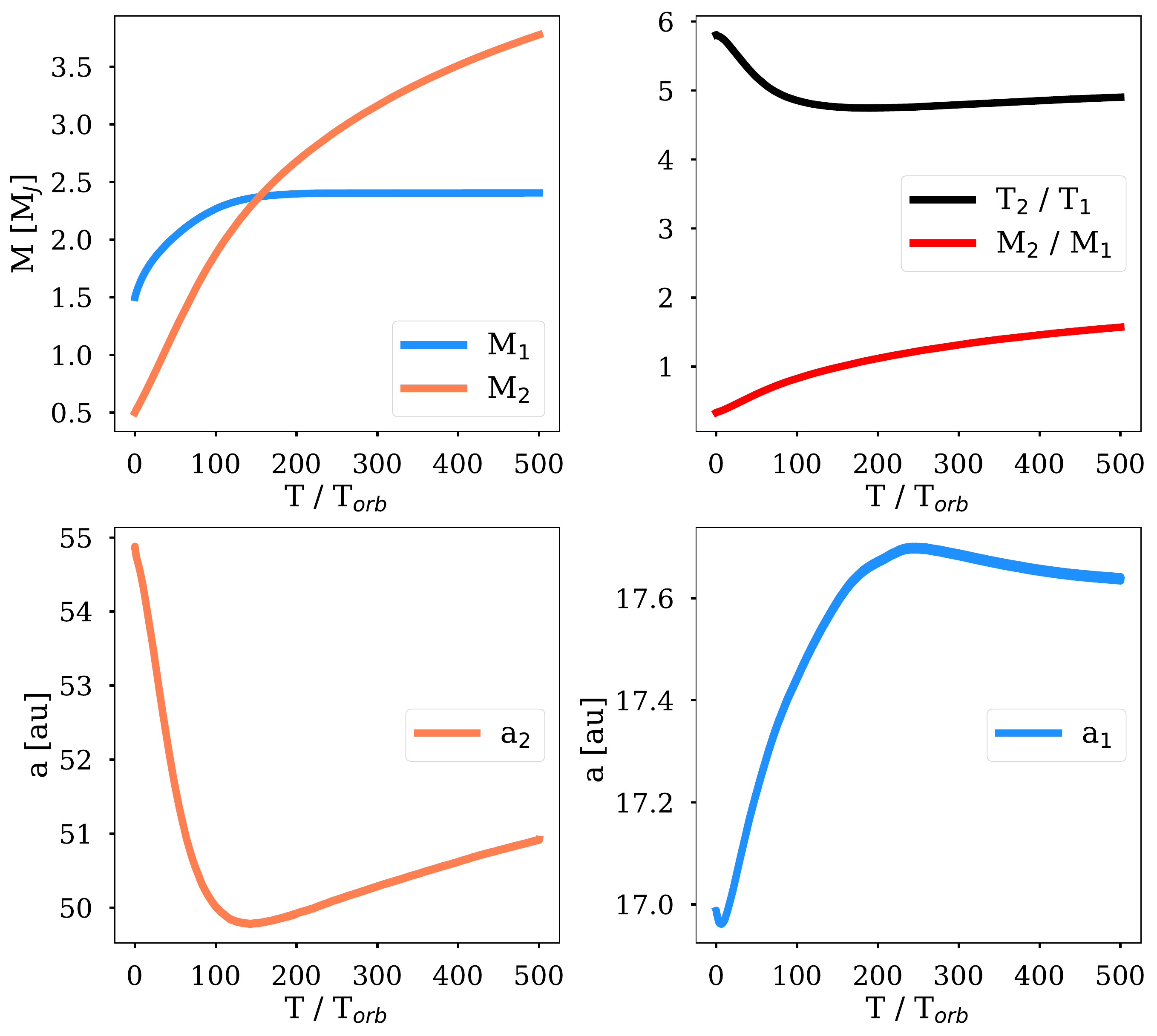}
\caption{Dynamics of the two planets. Top right: the masses of the planets; top left: the ratio between the outer and inner periods and masses; the orbital semi-major axis  of the outer (lower left) and inner planets (lower right) as a function of the normalised time. } 
\label{fig:2}
\end{minipage}
\hspace{.35cm}%
\begin{minipage}[c]{0.45\linewidth}
\includegraphics[width=1\linewidth]{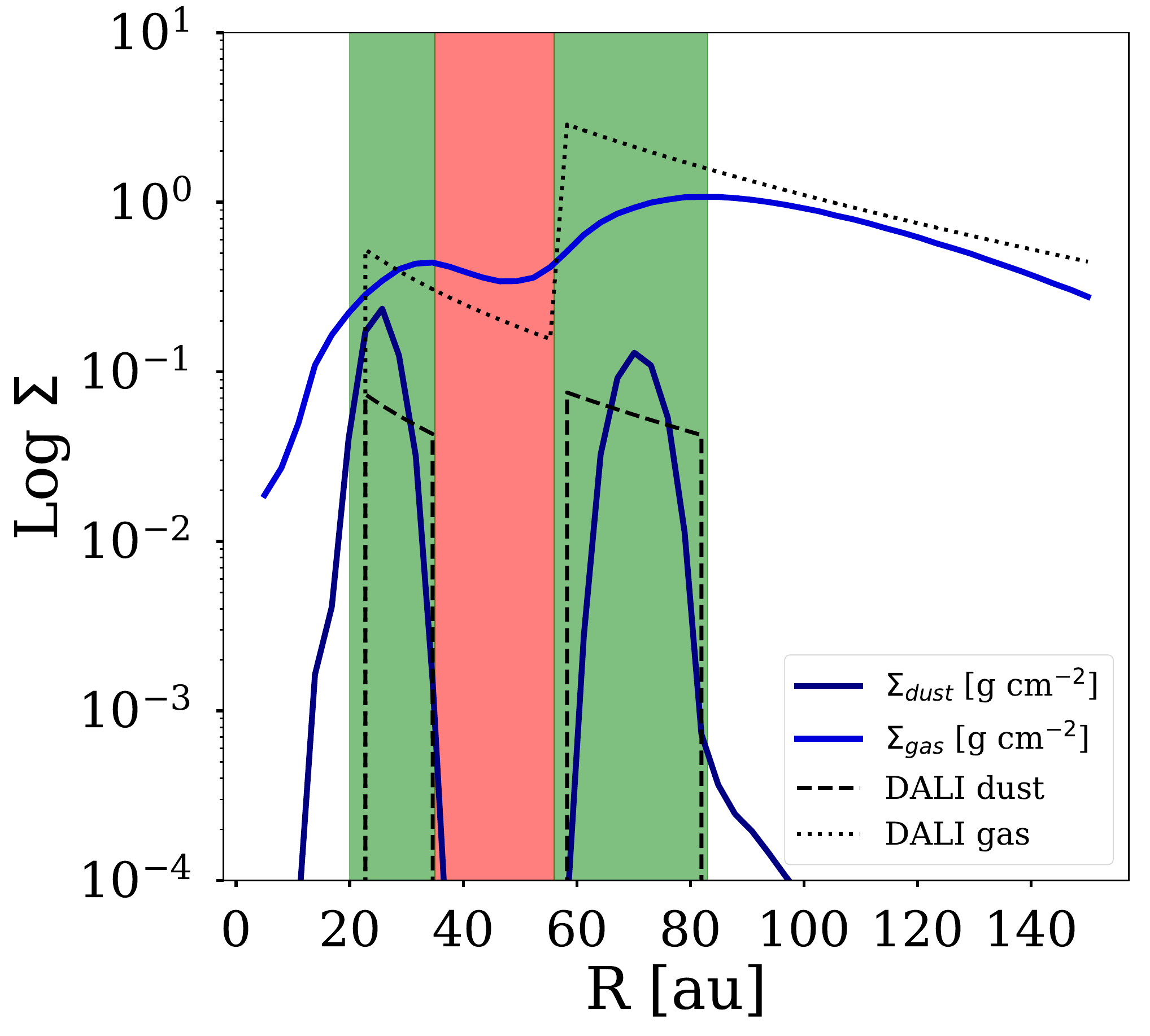}
\caption{Comparison between the $\Sigma_{gas}$ and $\Sigma_{dust}$ profiles after 160 $T_{\rm{orb}}$ and the update DALI model of FD17. The green and red regions are the observed locations of the rings and the gap respectively.}
\end{minipage}
\label{fig:3}
\end{figure}

Fig. \ref{fig:1} shows the temporal evolution of the dust and gas surface density profiles, $\Sigma_{dust}$ and $\Sigma_{gas}$, in our simulation. Going from black to red, $\Sigma_{dust}$ and $\Sigma_{gas}$ are displayed as a function of time, that we recall is normalised to the orbital period of the outer planet at the initial position (55 au, $	\rm{T}_{orb}\sim 320$~yr). 

 The inner planet carves a cavity in the dust and in the gas, with a depletion factor of $\sim 10^{-4}$ and $\sim 10^{-1}$ respectively inwards of its orbit. The outer planet first migrates (see Fig. \ref{fig:2}) while accreting the material, then opens a gap in dust and in gas (with depletion factor $\sim 10^{-4}$ and $\sim 10^{-1}$ respectively in dust and gas) at his 'equilibrium' location. The dust ring produced between the two planets is preserved for at least $\sim$ 500 orbits, i.e for over $\sim 2\times 10^5$~yr. At the same time, the gas is accreted on the star and on the planets.
The density profiles of the gas and dust in our simulation are due to a combinatio of factors. For the gas, the ``cavity'' is produced by the torques exerted by the outer planet onto the disc, that prevents the gas outside the outer planet to refill the inner regions, that are emptied by viscous accretion. For the dust, the two prominent rings are created by the usual trapping at pressure maxima (indeed, we have verified that the ring location coincides exactly with pressure maxima).
This generates a gas cavity with a width of $\sim $ 50~au, lowering the gas surface density and increasing the value of the Stokes number that quickly becomes larger than unity, helping the dust trapping due to the pressure gradient in preventing the dust migration and thus preserving the dust ring.

The outer dust ring is located outside the outer planet and  shrinks for dust migration as the disc evolves  for the first $\sim 150$ T$_{\rm{orb}}$. Later on ($t > $ 200 $\rm{T}_{\rm{orb}}$), the dust ring radius increases with time and the ring moves outward as the maximum of the gas pressure moves outward, in agreement with a dust trapping due to pressure bump scenario \citep{Pinilla2012}.
As time passes by, the width of the gap grows as the mass of the second planet grows, as expected from \citealt{Crida2006} and \citealt{Dipierro2017}.
As visible in Fig \ref{fig:1}, during the process (at $\sim 100-200$ orbits) the inner edge of the inner ring spreads inwards, making the dust surface density profile skewed (compare the green and cyan lines with the blue and red lines, that refers to early and later time respectively), as found also in PS19.

Fig. \ref{fig:2} shows the dynamics of the planets. In the first $\sim 50$ orbits the system relaxes out of the initial conditions: the outer planet rapidly migrates to $\sim 50$~au (lower left panel), while the inner planet remains at $\sim 17$~au, slowly migrating outward (lower right panel) . This leads the orbits of the two objects to become in a 5:1 period resonance (upper right panel).
Fig. \ref{fig:2} also shows the masses of the two planets as a function of time (upper left panel). After the first $\sim 100~\rm{T}_{orb}$ the inner planet has already accreted almost all of the available mass, reaching the value of $\sim 2.5~\rm{M}_J$, while the outer planet continues to accrete mass from the external ($\rm{R} > 60$~au) reservoir of gas, overcoming the mass of the inner planet (initially $\rm{M}_{\rm{1}} > \rm{M}_{\rm{2}}$, while at the end of the simulation $\rm{M}_{\rm{2}} > \rm{M}_{\rm{1}}$). This leads to a small change in the position of the two planets, the inner actually migrating outwards.
We underline that, while the outer planet is located almost in the middle of the outer gap, our simulation shows that the inner planet is actually located at the inner edge of the ring. The outward migration of the planet generates the skewness of the ring shown in Fig. \ref{fig:1}: it is actually dust accreted on the planet via a dusty circumplanetary disc.

Our best agreement with the results of the updated version of
 thermo-chemical model
of FD17 is obtained after $160~\rm{T}_{orb} = 5 \times 10^{4}$~yr and it is shown in Fig. \ref{fig:3}. The total mass of the disc is $\sim 6\times 10^{-3}$~M$_\odot$ and the dust mass is $\sim 0.9\times 10^{-4} ~\rm{M}_\odot$. The inner planet is located at $\sim 17.4$~au and the outer at $\sim50$~au, with almost equal masses, $\sim 2.4$~M$_J$.  

\begin{figure}
\centering
\includegraphics[width=0.67\linewidth]{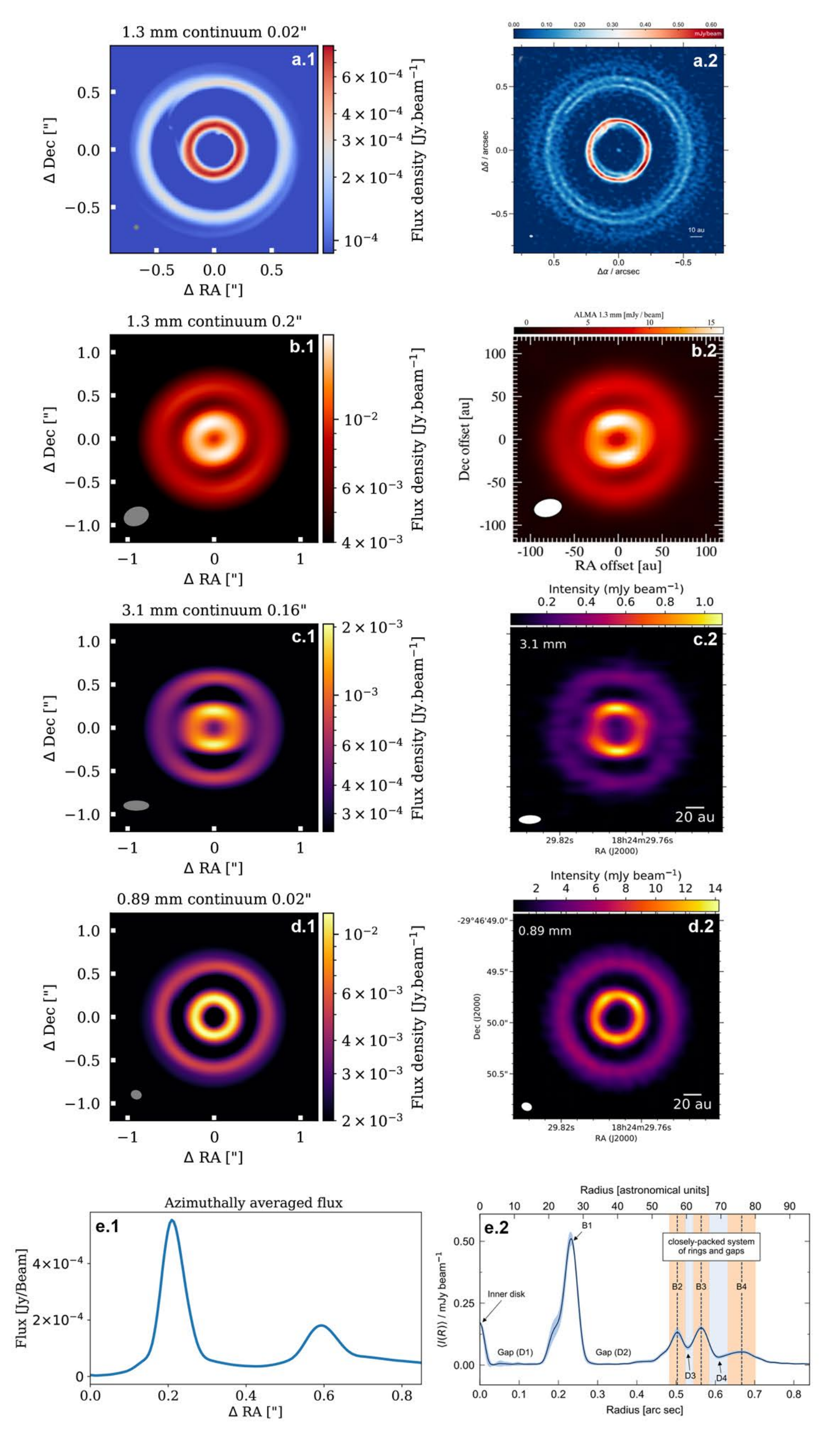}
\caption{Comparison between simulated observations of our disc model and ALMA observations. a.1,b.1,c.1,d.1, e.1: model, a.2,b.2,c.2,d.2,e.2: observations from PS19 (1.3 mm continuum emission), FD17 (1.3 mm), ME18 (3.1 and 0.89 mm) respectively . The grey filled ellipse in the lower left corner indicates the simulated beam size.}
\label{fig:4}
\end{figure}

Fig. \ref{fig:4} displays the synthetic observations of our simulation. First and second line panels show respectively our model continuum emission at 1.3 mm convolved with a 0.02'' (image a.1) and 0.2'' beam (image b.1), in good agreement with the observed fluxes of PS19 and FD17 (images a.2 and b.2 respectively). 
Third and fourth line show 3.1 mm at 0.16'' resolution (image c.1) and 0.890 mm at 0.02'' (image d.1) simulated observations of our disc, in good agreement with ME18 observations (c.2 and d.2 respectively).
Our continuum emission images has been obtained supposing a chemical composition for the dust grains of 70$\%$ silicates and 30$\%$ carbonaceus material, with a porosity level of 10$\%$. 
Panels e.1 and e.2 show the azimuthally averaged profile of our model 
and PS19 respectively. The ring profile is well reproduced and the value of the flux is consistent with the observation. 

\begin{figure}[ht!]
\centering
\includegraphics[width=0.6\linewidth]{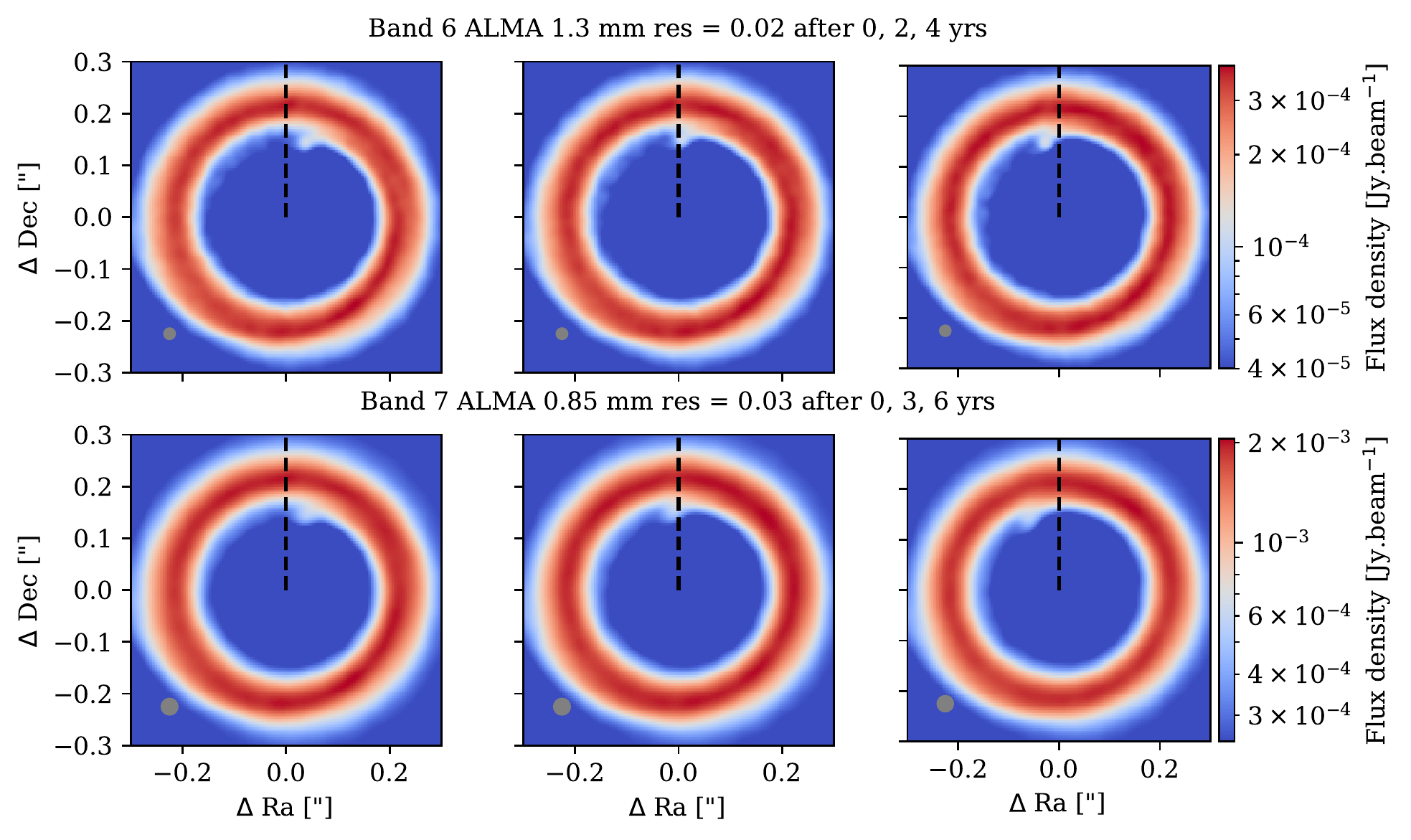}
\caption{ Zoom of the inner part of the model. The inner planet is located close to the inner edge of the ring, generating an accretion feature visible in our model and 
expected to move covering $\sim 5$ degrees in $\sim 1$ yr. Top: ALMA band 6 observations at three different times separated by tree years each.  Bottom: ALMA band 7 observations at three different times separated by two years each. The grey filled ellipse in the lower left corner indicates the simulated beam size.}
\label{fig:5}
\end{figure}

An interesting feature of our simulation, as mentioned above, is the creation of a distinct dust feature at the inner edge of the inner ring, probably due to dusty material accreted from the disc disposed around the inner planet in a circumplanetary disc. This is clearly visible also in our simulated ALMA observations in Fig. \ref{fig:4}. Such a feature might have been already observed in the high-resolution of PS19 (see the radially extended feature in their Fig. \ref{fig:2}, at an angle of $\sim$ 40 degrees). In order to confirm this hypothesis, one might check the evolution of the feature with time, due to the Keplerian rotation of the inner planet, which has a period of $\sim 60$~yr assuming a distance of $\sim18$~au. Fig. \ref{fig:5} shows a zoom of the inner part of our model. The top panel refers to simulated ALMA band 6 observations, while the lower panel refers to ALMA band 7. From left to right the plot refer to three different times in the simulations, separated by three years each for the band 6 and two years each for the band 7. Future observations of this system might clearly detect such shift in the azimuthal location of the feature, as the planet is expected to rotate by a 5 degree angle in one year.  

\section{Discussion} \label{sec:discussion}
The presence of two giant planets shapes the disc, explaining the presence of the two rings in the dust and the cavity in the gas. 
The dust gap quickly opens (after $40-60$  T$_{\rm{orb}}$) due to pure gravitational effects \citep{Crida2006} and dust radial drift \citep{Dipierro2017}. The gas gap opens at slower rate due to the effects of the interaction between the planets and the disc and the viscous evolution of the disc itself.
The presence of the planets creates two pressure bumps, where the dust rings are trapped.
After $\sim 100-120$ $\rm{T}_{\rm{orb}}$, due to the high masses of the planets the inner part (with the gas cavity and the dust ring) starts to be disconnected from the outer part, that includes the outer dust ring at $\sim 50-70$~au and a reservoire in gas that extends for radii larger than $\sim 150$~au. 

The outer planet is still accreting material from the gas reservoire, increasing its mass. This interaction between the outer planet and the disc prevents also the refilling of the inner gas gap. Hovever, the star is also accreting material from the inner part of the disc due to viscous accretion. As a consequence, the gas surface density decreases with time, creating a cavity for R $< 50$ au. This reduces the effect of the pressure bump in the inner part of the disc, but increases the value of the Stokes number. The global effect is that the inner dust ring is preserved in time.
We do not reproduce the outer ring splitting including a third planet, as this is out of the purpose of the present study, but we note that the interpretation of PS19 (a third planet) is a likely possibility. 

The longevity of this configuration 
allows the formation and migration of outer protoplanets as found in PS19.
The dust trapped in the inner region can grow in size, explaining what observed in the 3.0 mm observation of ME18.
Indeed, we also obtain the absence of dust grains inside the cavity and the gap with a surface density reduction (compared to the assumed initial structure) of $\sim 30-40$ and a gas to dust ratio of $\sim 1$ as found in FD17 and ME18.

The planetary system displayed in our simulation, with two giant planets in orbital resonances, is a common feature of the observed exoplanetary systems and present also in our Solar System. Our simulation also seems to point out that giant planets tend to reach orbital resonances while still
embedded in the parental disc. 
This fact, combined with the gravitational interaction between them, led in our configuration to a small outward migration of the inner planet, that orbits close to the inner edge of the innermost dust ring and not in the center of the cavity as expected. This behaviour is not dependent on the initial condition of the planets, free to migrate.
 
\section{Conclusions} \label{sec:conclusions}
We run hydrodynamic simulations and reproduce long lived dust rings between two giant planets, a scenario that can model the observations of the inner region of the protoplanetary disc HD~169142. 

We included two giant planets free to migrate, initially not in orbital resonant positions, obtaining our best fit with masses of $\sim 2.4$ M$_J$ at $17.4$ and $50$~au, in a resonant locking of 5:1. The total mass of the disc is $\sim 6\times 10^{-3}$~M$_\odot$ and the dust mass is $\sim 0.9\times 10^{-4} ~\rm{M}_\odot$.

Our result shows that is possible to form and preserve a dust rings between the orbits of two giant planets for at least $\sim 2\times 10^5$ yr, a significant fraction of the disc lifetime.
The dust is initially trapped in the pressure maximum between the two planets. As the gas surface density decreases inside the gas gap, the value of the Stokes number increases to values larger than unity, helping the pressure bump in preserving the dust to migrate.

The combined effect between the planets-disc interaction and the gas accretion on the planets and on the star leads to the shaping of a decreased (at least one order of magniture) region in the gas surface density profile (R $< 50 $ au), as expected from the CO lines observations of FD17.
The fact that our code allows planet to accrete and migrate reaching minimum energy configurations and resonances is crucial for obtaining this results.

While the outer planet is located at the centre of the gap, the inner one is close to the inner edge of the dust ring and can eventually accrete material from it, generating a feature that might be observable and can explain the ALMA 1.3 mm observation of SP19. Further observations with ALMA are crucial to understand if this feature is really due to a protoplanet accreting in the cavity.

This configuration is relevant also to explain the dynamics of observed cases as PDS 70 \citep{Keppler2019}, where the outer planet is located ad the edge of the dust ring and is actively accreting material. However, in the case of PDS70 the inner planet has a mass that is larger than the outer one, a situation that is opposite to our scenario.
\section{Acknowledgments}
The authors want to thank the referee for constructive comments that improved this manuscript. This work and CT are supported
by the PRIN-INAF 2016 The Cradle of Life - GENESIS-SKA (General Conditions in Early Planetary Systems for the rise of life with SKA).
CT, DF and LT acknowledge financial support provided by the Italian Ministry of Education, Universities and Research. CT and LT through the grant Progetti Premiali 2012 – iALMA (CUP C52I13000140001), DF from project SIR (RBSI14ZRHR); LT aknowledges the Deutsche Forschungsgemeinschaft (DFG, German Research Foundation) - Ref no. FOR 2634/1ER685/11-1 and the DFG cluster of excellence ORIGINS (www.origins-cluster.de).
GL and LT received funding from the EU Horizon 2020 research and
innovation programme, Marie Sklodowska-Curie
grant agreement 823823 (Dustbusters RISE project). 
CP acknowledges funding from the Australian
Research Council via FT170100040 and
DP180104235 and from ANR of France (ANR-16-CE31-0013).
This paper makes use of the following ALMA data: ADS/JAO.ALMA 2013.1.00592.S. ALMA is a partnership of ESO (representing its member states), NSF (USA) and NINS (Japan), together with NRC (Canada), MOST and ASIAA (Taiwan), and KASI (Republic of Korea), in cooperation with the Republic of Chile. The Joint ALMA Observatory is operated by ESO, AUI/NRAO and NAOJ.


\begin{thebibliography}{}

\bibitem[\protect\citeauthoryear{ALMA partnership}{2015}]{ALMA2015}
ALMA Partnership, 2015, ApJL, 808, 1,L3 

\bibitem[\protect\citeauthoryear{Alexander et al.}{2014}]{Alexander2014}
Alexander R. et al., 2014, Protostars and Planets VI 

\bibitem[\protect\citeauthoryear{Andrews et al.}{2018}]{Andrews2018}
Andrews S. M. et al., 2018, ApJ, 869, L41

\bibitem[\protect\citeauthoryear{Ballabio et al.}{2018}]{Ballabio2018}
Ballabio G. et al., 2018, MNRAS, 477, 2766

\bibitem[\protect\citeauthoryear{Bate et al.}{1995}]{Bate1995}
Bate M. R. et al., 1995, MNRAS, 277, 362

\bibitem[\protect\citeauthoryear{Benisty et al.}{2015}]{Benisty2015}
Benisty M., et al. 2015, A$\&$A, 578, L6

\bibitem[\protect\citeauthoryear{Bertrang et al.}{2018}]{Bertrang2018}
Bertrang, G. H.-M. et al. 2018, MNRAS, 474, 5105.

\bibitem[\protect\citeauthoryear{Birnstiel et al.}{2015}]{Birnstiel2015}
Birnstiel T. et al., 2015, ApJ, 813, L14

\bibitem[\protect\citeauthoryear{Brauer et al.}{2008}]{Brauer2008}
Brauer, F. et al., 2008, A$\&$A,
480, 859

\bibitem[\protect\citeauthoryear{Crida et al.}{2006}]{Crida2006}
Crida et al., Icarus 181, 587

\bibitem[\protect\citeauthoryear{Dipierro et al.}{2015}]{Dipierro2015}
Dipierro G. et al. 2015, MNRAS, 453, L73

\bibitem[\protect\citeauthoryear{Dipierro $\&$ Laibe}{2017}]{Dipierro2017}
Dipierro G. $\&$ Laibe, G., 2017, MNRAS, 479, 2 

\bibitem[\protect\citeauthoryear{Draine}{2003}]{Draine2003}
Draine B. T., 2003, ApJ, 598, 1017

\bibitem[\protect\citeauthoryear{Fedele et al.}{2017}]{Fedele2017}
Fedele, D. et al., 2017, A$\&$A, 600, A72.

\bibitem[\protect\citeauthoryear{Fressin et al.}{2013}]{Fressin2013}
Fressin F.  et al., 2013, ApJ, 766, 81

\bibitem[\protect\citeauthoryear{Gaia Collaboration}{2016}]{Gaia2016}
Gaia Collaboration 2016, A$\&$A, 595, A2

\bibitem[\protect\citeauthoryear{Gonzalez et al.}{2015}]{Gonzalez2015}
Gonzalez J.F., et al., 2015, MNRAS, 454, L36

\bibitem[\protect\citeauthoryear{Gratton et al.}{2019}]{Gratton2019}
Gratton, R. et al. 2019, arXiv e-prints,
arXiv:1901.06555

\bibitem[\protect\citeauthoryear{Isella et al. }{2019}]{Isella2019}
Isella A. et al, 2019, arXiv e-prints, arXiv:1906.06308

\bibitem[\protect\citeauthoryear{Keppler et al.}{2018}]{Keppler2019}
Keppler, M. et al. 2018, arXiv:1806.11568

\bibitem[\protect\citeauthoryear{Honda et al.}{2012}]{Honda2012}
Honda, M. et al. 2012, ApJ, 752, 143.

\bibitem[\protect\citeauthoryear{Huang et al.}{2018}]{Huang2018}
Huang, J. et al, 2018, ApJ, 852, 2.

\bibitem[\protect\citeauthoryear{Laibe $\&$ Price}{2014}]{Laibe2014}
Laibe $\&$ Price 2014, MNRAS, 440, 3, 21

\bibitem[\protect\citeauthoryear{Ligi et al.}{2018}]{Ligi2018}
Ligi, R. et al. 2018, MNRAS, 473, 1774

\bibitem[\protect\citeauthoryear{Lin $\&$ Papaloizou}{1986}]{LP1986}
Lin D. N. C., Papaloizou J., 1986, ApJ, 307, 395

\bibitem[\protect\citeauthoryear{Liu et al.}{2019}]{Liu2019}
Liu Y., et al., 2019, A$\&$A, 622, A75

\bibitem[\protect\citeauthoryear{Lodato et al.}{2010}]{Lodato2010}
Lodato G., Price D. J., 2010, MNRAS, 405, 1212

\bibitem[\protect\citeauthoryear{Lodato et al.}{2019}]{Lodato2019}
Lodato, G. et al. 2019, MNRAS, 486, 453

\bibitem[\protect\citeauthoryear{Long et al.}{2018}]{Long2018}
Long F., et al., 2018, ApJ, 869, 17

\bibitem[\protect\citeauthoryear{Macias et al}{2018}]{Macias2018}
Macias, E. et al. 2018, ApJ, 865, 37.

\bibitem[\protect\citeauthoryear{Mordasini et al.}{2016}]{Mordasini2016}
Mordasini, C. et al, 2016, APJ, 832, 41

\bibitem[\protect\citeauthoryear{Osorio et al.}{2014}]{Osorio2014}
Osorio, M. et al. 2014, ApJ, 791, L36

\bibitem[\protect\citeauthoryear{Pani{\'c} et al.}{2008}]{Panic2008}
Pani{\'c}, O. et al. , C. 2008, A$\&$A, 491, 219

\bibitem[\protect\citeauthoryear{Pinte et al.}{2006}]{Pinte2006}
Pinte, C. et al., 2006, A$\&$A, 459, 3, 797

\bibitem[\protect\citeauthoryear{Perez et al.}{2019}]{Perez2019}
P{\'e}rez, S. et al, arXiv e-prints, arXiv:1902.05143

\bibitem[\protect\citeauthoryear{Pinte}{2009}]{Pinte2009}
Pinte C. et al, 2009, A$\&$A, 498, 967

\bibitem[\protect\citeauthoryear{Pinte}{2018}]{Pinte2019}
Pinte, C., Price, D. J., Menard, F., et al. 2018, arXiv:1805.10293

\bibitem[\protect\citeauthoryear{Pinilla et al.}{2012}]{Pinilla2012}
Pinilla P. et al, 2012, A$\&$A, 545, A81

\bibitem[\protect\citeauthoryear{Pinilla et al.}{2016}]{Pinilla2016}
Pinilla P. et al, 2016, A$\&$A, 596, A81

\bibitem[\protect\citeauthoryear{Pohl et al.}{2017}]{Pohl2017}
Pohl, A.  et al., 2017, ApJ, 850, 52.

\bibitem[\protect\citeauthoryear{Price et al.}{2018}]{Price2017}
Price D. J. et al., 2018, PASA, 35, e031

\bibitem[\protect\citeauthoryear{Reggiani et al.}{2014}]{Reggiani2014}
Reggiani, M.  et al. 2014, ApJ, 792,
L23

\bibitem[\protect\citeauthoryear{Sakura $\&$ Sunyaev}{1973}]{SS1973}
Shakura N. I., Sunyaev R. A., 1973, A$\&$A, 24, 337

\bibitem[\protect\citeauthoryear{Siess et al.}{2000}]{Siess2000}
Siess L. et al., 2000, A$\&$A, 358, 593

\bibitem[\protect\citeauthoryear{Suriano et al.}{2018}]{Suriano2018}
Suriano, S. S. et al.,
2018, MNRAS, 477, 1239, doi: 10.1093/mnras/sty717

\bibitem[\protect\citeauthoryear{Testi et al.}{2015}]{Testi2015}
Testi et al., 2015, ApJ, 812, 2

\bibitem[\protect\citeauthoryear{Wagner et al.}{2018}]{Wagner2018}
Wagner, K.  et al. 2018, ApJL, 863, L8

\end{thebibliography}
\end{document}